\documentclass{article} 
\usepackage{iclr2024_conference,times}

\usepackage{amsmath,amsfonts,bm}

\def\eqref#1{equation~\ref{#1}}

\def\1{\bm{1}}

\DeclareMathAlphabet{\mathsfit}{\encodingdefault}{\sfdefault}{m}{sl}
\SetMathAlphabet{\mathsfit}{bold}{\encodingdefault}{\sfdefault}{bx}{n}

\usepackage{hyperref}
\usepackage{url}

\usepackage{amsmath}  
\usepackage{bm}       
\usepackage{graphicx}
\usepackage{subcaption}
\usepackage{authblk}

\title{From Noise to Signal: Unveiling Treatment Effects from  Digital Health Data through Pharmacology-Informed Neural-SDE}

\author[1, 2, +]{Samira Pakravan}  
\author[1, 3, +]{Nikolaos Evangelou} 
\author[4]{Maxime Usdin} 
\author[1, *]{Logan Brooks}
\author[1, *]{James Lu}

\affil[1]{\footnotesize Clinical Pharmacology, Genentech, South San Francisco, CA 94080, USA}
\affil[2]{\footnotesize Department of Mechanical Engineering, University of California, Santa Barbara, CA, USA}
\affil[3]{\footnotesize Department of Chemical and Biomolecular Engineering Johns Hopkins University Baltimore, MD, USA}
\affil[4]{\footnotesize Computational Sciences, Genentech, South San Francisco, CA 94080, USA}
\affil[+]{\footnotesize Equal contribution}
\affil[*]{\footnotesize Corresponding authors ( brooksl2@gene.com, lu.james@gene.com)}

 \iclrfinalcopy 

\begin{document}

\maketitle

\begin{abstract}
Digital health technologies (DHT), such as wearable devices, provide personalized, continuous, and real-time  monitoring of patient. These technologies are contributing to the development of novel therapies and personalized medicine. Gaining insight from these technologies requires appropriate modeling techniques to capture clinically-relevant changes in disease state. The 
data generated from these devices is 
characterized by being
stochastic in nature, may have missing elements, and exhibits considerable inter-individual variability - thereby making it difficult to analyze using traditional longitudinal modeling techniques. We present a novel pharmacology-informed neural stochastic differential equation (SDE) model capable of addressing these challenges. Using synthetic data, we demonstrate that our approach is effective in identifying treatment effects and learning causal relationships from stochastic data, thereby enabling counterfactual simulation.  
\end{abstract}

\section{Introduction}
The rise of digital health technologies (DHT)  including wearable devices such as smart watch and patch based physiological sensors has opened new possibilities for continuous patient monitoring  \cite{friend2023wearable} and enables generation of time-series data at an unprecedented temporal resolution and duration, thereby offering the potential to generate new clinical measures and insights \cite{berisha2021digital}. Furthermore, recent examples have shown the clinical value in modeling both the longitudinal trends as well as the stochastistity in digital health (DH) data \cite{leander2022stochastic}.

Stochastic differential equations (SDEs) have been developed to describe various phenomena that exhibits random fluctuations \cite{fagin2023latent}, including in biological and biomedical applications \cite{mei2013enisi, tajmirriahi2021modeling}. In the context of DH, the interplay between physiology and the measurement device is likely far too complex for one to theoretically derive the equations underlying the link between disease status and DH data from first principles. Instead, we propose to learn the underlying dynamical system directly from data, with the help of neural-SDE \cite{evangelou2022learning, dietrich2023learning}.

Here, we develop a \textit{pharmacology-informed} \cite{lu2021deep, laurie2023explainable} neural-SDE that:
\begin{itemize}
    \item learns the underlying dynamical system from a patient population, while introducing patient-dependent parameters that enables the characterization of patient-to-patient variability;
    \item incorporates the causality between pharmacokinetics (PK) and pharmacodynamics (PD);
    \item enables counterfactual simulations to describe drug effects at the individual patient level.
\end{itemize}
We demonstrate the effectiveness of the proposed model using synthetic data.

\section{Methods}

\subsection{Neural-SDE Model}
We assume that the longitudinal data are modelled by a system of equations of the form,

\begin{align}
    \label{eqn:ode}
    dc_t &= f(c_t)dt \\
    \label{eqn:sde}
    dx_t &= \nu(x_t,c_t,\bm{p})dt + \sigma(x_t,c_t,\bm{p})dW_t
\end{align}

where Equation \ref{eqn:ode} represents a known Ordinary Differential Equation (ODE) model with $f(\cdot)$ being the vector field for PK that governs the drug concentration, $c_t \in \mathbb{R}$ , and   
where the drift and diffusion terms (i.e., $\nu(x_t,c_t,\bm{p})$ and $\sigma(x_t,c_t,\bm{p})$ respectively) are described by neural networks.
%
%
We work under the hypothesis that the drift and diffusivity terms of the \textit{effective} SDE, 
are dependent on the state ($x_t \in \mathbb{R}$) as well as the drug concentration $c_t$. Additionally, while the underlying equations are the same for all patients, the model includes  a \textit{latent} patient-dependent parameter vector $\bm{p}$ that describes the patient-to-patient variability. This \textit{latent} parameter $\bm{p}$ is discovered in a data-driven way based on the work of \citet{lu2021deep}, which we elaborate below. 

While the available data are in the form of trajectories, we transform them to snapshots $\mathcal{D}$ in a manner analogous to that done in \cite{dietrich2023learning}. In particular, each snapshot $\mathcal{D}^i$, uniquely identified by the index $i$, takes the form $\mathcal{D}^i = \{x_1^i,x_0^i,\Delta t,c_1^i,\bm{p}^{i,j}\}$, where $x^i_1$ is the evolution of the state variable $x_t$ after a time step $\Delta t$ given the initial condition $x_0^i$; $\bm{p}^{i,j}$ is the \textit{latent} parameter for the $j$th patient. Note that we utilize the concentration at $c_1$ and not at $c_0$ following the (symplectic) Euler-Maryama scheme discussed in \citet{dietrich2023learning}. The concentration $c_t$ and the patient dependent parameter $\bm{p}$ enter into the overall architecture as inputs based on  \citet{dietrich2023learning,evangelou2022learning}.

The construction of the loss function (based on \citet{dietrich2023learning}) is derived from the numerical integration scheme (symplectic) Euler-Maruyama. The numerical approximation of Equations \ref{eqn:ode} and \ref{eqn:sde} results in: 
\begin{align}    
    & c^i_1 = c^i_0 + f(c^i_0)\Delta t \\
    & x^i_1 = x^i_0 + \nu(x^i_0,c^i_1,\bm{p}^{i,j})\Delta t + \sigma(x^i_0,c^i_1,\bm{p}^{i,j})\delta W_0,   
    \label{eqn:Symplectic_Euler_Maryama}
\end{align}

where $\delta W_0$ is normally distributed around zero and $\Delta t$ is a variable timestep. The drift and diffusivity terms are approximated by two networks $\nu_{\theta}$ and $\sigma_{\theta}$, under the assumption that $x_1$ is drawn from a normal distribution of the form,
\begin{equation}
    x_1^i \sim \mathcal{N}(x_0^i + \nu_\theta(x_0^i,c_1^i,\bm{p}^{i,j})\Delta t,\sigma_{\theta}(x_0^i,c_1^i,\bm{p}^{i,j})^2\Delta t).
    \label{eqn:Normal_Distribution}
\end{equation}


With the assumed mean and variance in Equation \ref{eqn:Normal_Distribution} for the drift and diffusivity, we can compute the logarithm of the resulting normal distribution and derive the following loss function that maximizes the likelihood:


\begin{equation}
\label{eqn:Loss_Function}
    \mathcal{L}(\theta|x_0^i,x_1^i,\Delta t) := \frac{(x_1^i - x_0^i - \nu_{\theta}(x_0^i,c_1^i, \bm{p}^{i,j}))^2}{\Delta t\sigma_\theta(x_0^{i},c_1^{i},\bm{p}^{i,j})^2} + \text{log}|\Delta 
 t\sigma(x_0^i,c_1^i,\bm{p}^{i,j})^2|.
\end{equation}

It should be noted that the Neural-SDE framework by  \citet{dietrich2023learning} is also capable of handling varying time steps $\Delta t$. 

\begin{figure}[ht]
\centering
\includegraphics[width=0.6\textwidth]{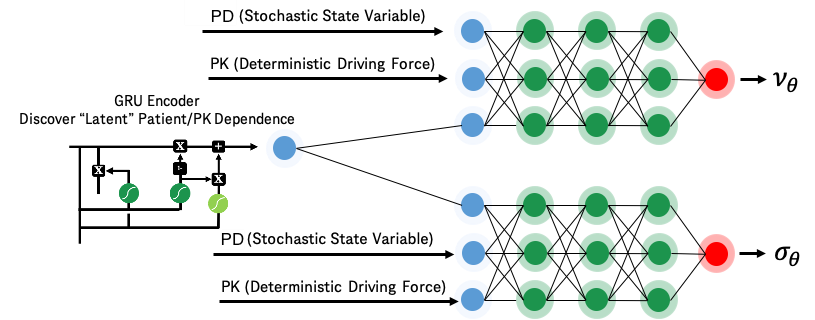}
\caption{The Neural-SDE architecture including the GRU encoder.}
\label{fig:Neural_SDE_architecture}
\end{figure}

The Neural-SDE architecture consists of two network components for the drift and diffusion models. In our work, the drift network consists of 4 layers where each layer has 64 neurons each followed by \texttt{ELU} activation function. The diffusion network consists of 3 layers with 32 neurons, the first two layers are followed by \texttt{ELU} activation function and the output layer is followed by \texttt{softplus} activation function. A schematic of the Neural-SDE architecture is shown in Figure \ref{fig:Neural_SDE_architecture}.

\subsection{Latent Patient Descriptors - GRU Encoder}


Our approach to learning the Neural-SDE from data across the patient population is to identify a set of dynamical equations  that holds across all patients, as well as patient-specific descriptors (or embedding) that characterize patient-to-patient variability \cite{laurie2023explainable}.
In our approach, those patient-specific descriptors are discovered in a data-driven manner, based on the work of  \citet{lu2021deep}: a Gated Recurrent Unit (GRU) encoder was used to discover the latent parameter $\bm{p}$, with longitudinal data provided in a tabular form as an input. More specifically, the input data entering the encoder consist of variable number of rows for each patient and the following four columns: (1) the absolute time; (2) the time after dose; (3) the stochastic PD data (4) the deterministic PK data. 

Each tabular input was \textit{padded} and \textit{masking} was applied in order to handle the variable time points. The GRU encoder has 128 hidden states and is connected to a Multilayer Perceptron (MLP) consisting of 2 layers, each with 128 neurons, both followed by \texttt{ELU} activation function. The output of MLP is the \textit{latent} parameter $\bm{p}$ that enters the Neural-SDE architecture. An end-to-end training was implemented by using the loss function given by Equation \ref{eqn:Loss_Function}.

\subsection{Dataset}
To mimic clinical digital health measurements, synthetic data was simulated in which the PK serves as a deterministic driving input that causally influences a stochastically evolving PD.
Patient specific parameters were sampled from a log-normal distribution:  50 individual patient trajectories were sampled across 3 different dose levels (50 mg, 100 mg, 400 mg) for a total of 150 patient trajectories and 70:30 train-test split was used; further details are summarized in Appendix \ref{Appendix:data_gen}.

\section{Results}
\begin{figure}[ht]
\centering
\includegraphics[width=0.7\textwidth]{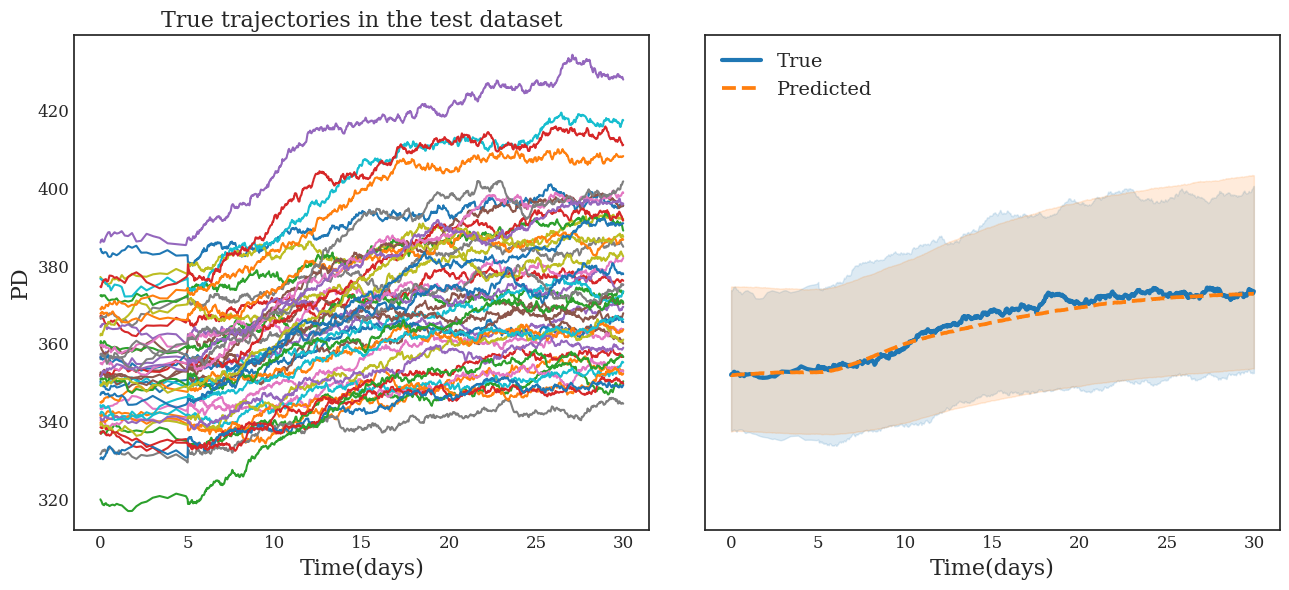}
\caption{Comparison of the true and predicted SDE trajectories in the test dataset.  Left panel: the colored lines represent the observed stochastic trajectories in the test data.  Right panel: blue line and shaded region represent the median and the $10^{th}$ to $90^{th}$ percentile respectively of the ground truth trajectories; similarly, the orange lines and shade region represent those from the model.}
\label{fig:All_test_trajectories}
\end{figure}

Figure \ref{fig:All_test_trajectories} demonstrates the model's ability to learn the underlying system's dynamics by comparing ``true'' (i.e., the underlying ground truth) SDE trajectories from the test dataset against the model predicted trajectories. For each patient in the test set, we sampled 250 trajectories to provide a robust representation of the predictive 
variability associated with the model. This result demonstrates the model's ability in replicating the complex dynamics of PD trajectories 
at the population level. 

\subsection{Dosing regimen analysis}
\begin{figure}[hbt!]
    \centering
    \begin{subfigure}{0.32\textwidth}
        \centering
        \includegraphics[width=\linewidth]{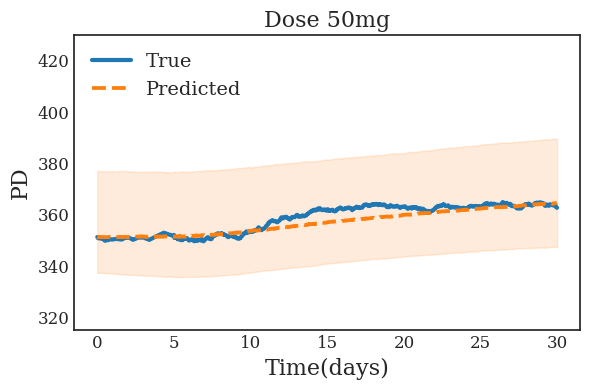}
        \label{fig:sub1}
    \end{subfigure}
    \hfill
    \begin{subfigure}{0.32\textwidth}
        \centering
        \includegraphics[width=\linewidth]{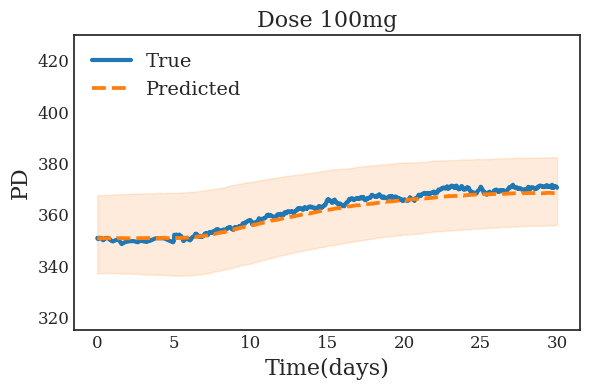}
        \label{fig:sub2}
    \end{subfigure}
   \hfill
    \begin{subfigure}{0.32\textwidth}
        \centering
        \includegraphics[width=\linewidth]{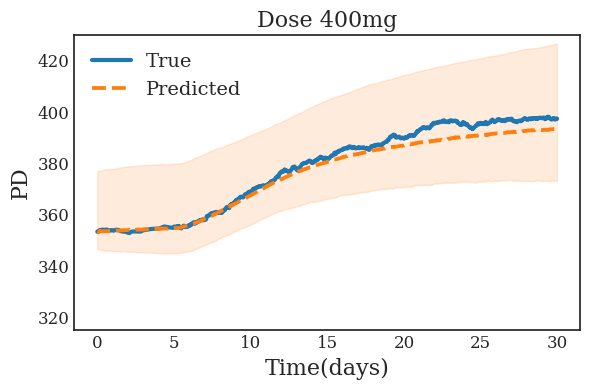}
        \label{fig:sub3}
    \end{subfigure}
    \caption{Comparison of the true and predicted SDE trajectories in the test datase for 50, 100 and 400 mg doses. Blue lines represent the median of the ground truth trajectories; orange dashed lines and shaded regions represent median and the $10^{th}$ to $90^{th}$ percentile of trajectories from the model.}
    \label{fig:Comparison_of_True_Predicted_regimen}
\end{figure}

To analyze the impact of different dosing regimens on PD, we consider three distinct simulated doses at $\rm 50~mg$, $\rm 100~mg$, and $\rm 400~mg$. For each patient from the test dataset, we sampled 250 SDE trajectories. Figure \ref{fig:Comparison_of_True_Predicted_regimen} shows the the model is qualitatively able to capture the true underlying dose response relationship.

\subsection{Patient-specific responses and counterfactual analysis}
\begin{figure}[hbt!]
    \centering
    \begin{subfigure}{0.32\textwidth}
        \centering
        \includegraphics[width=\textwidth]{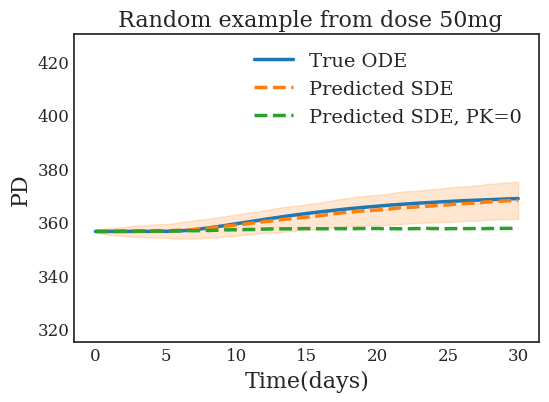}
        \label{fig:sub1}
    \end{subfigure}
    \hfill
    \begin{subfigure}{0.32\textwidth}
        \centering
        \includegraphics[width=\textwidth]{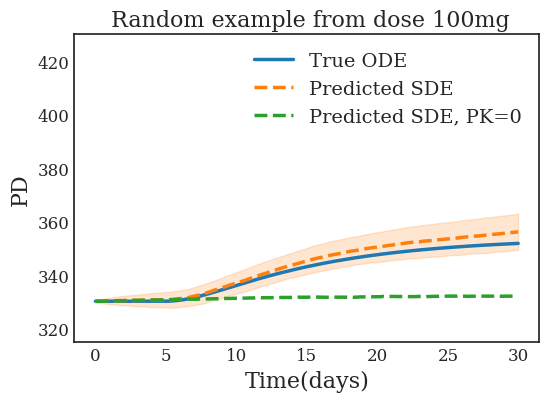}
        \label{fig:sub2}
    \end{subfigure}
    \hfill
    \begin{subfigure}{0.32\textwidth}
        \centering
        \includegraphics[width=\textwidth]{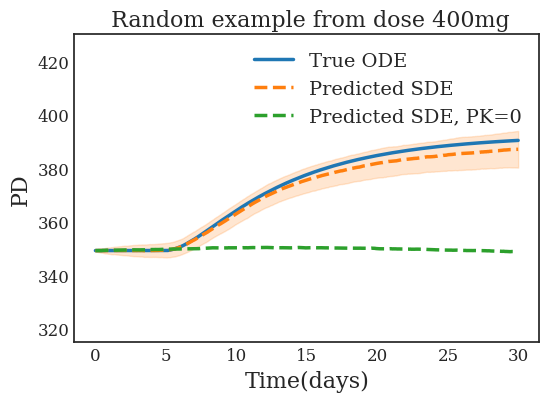}
        \label{fig:sub3}
    \end{subfigure}
    \caption{Patient-specific trajectories and counterfactual simulations. Each subplot represents a random patient from the respective dosages. The solid blue line represents the true drift; the orange dashed line and  shaded region represent the mean and mean $\rm \pm$ standard deviation (std) of 250 posterior samples; the green dashed lines represent counterfactual simulations assuming no dosing (\textit{i.e.}, $\rm PK=0$).}
    \label{fig:Patient-Specific_Counterfactual}
\end{figure}

Figure \ref{fig:Patient-Specific_Counterfactual} demonstrates the proposed methodology's ability to perform counterfactual analysis and identify individual treatment effects. To accomplish this, for each patient the drift and diffusivity terms were inferred from the trained model and 250 SDE trajectories were generated. The results demonstrate the model's ability to capture the underlying dynamics of the stochastic process for individual patients. This suggests that the GRU encoding strategy not only captures the population behaviors, but also successfully learns to differentiate amongst patients. Moreover, we demonstrate a what-if scenario: in the absence of PK, the model correctly predicts a lack of dynamical change in the modeled PD endpoint. This suggests our model is able to correctly identify the causal relationship between PK and PD. 

\section{CONCLUSION}
We proposed a pharmacology-informed neural-SDE architecture that is able learn the relationship between a deterministic PK and stochastic PD. Using synthetic data, the model correctly reproduces the underlying PK-PD relationship at the population level. Furthermore, the model enables the counterfactual simulation of  PD in the absence of the hypothetical drug - and in doing so, quantify the individual treatment effect.

\bibliography{iclr2024_conference}
\bibliographystyle{iclr2024_conference}

\appendix

\section{Appendix}
\subsection{Dataset Generation Details}
\label{Appendix:data_gen}
Synthetic training data was generated to represent a indirect response PK-PD model \citet{dayneka1993} by which PK acts causally to change the PD, with the additional modification that the observable PD variable is stochastic in nature. This system follows the general form of Equations \ref{eqn:ode} and \ref{eqn:sde}, 
%
with the following system of ODEs being specified for the term $f(c_{t}, \bm{p})dt$: 

\begin{align}
    \label{eqn:syn_data_ode}
    \frac{du_{1}}{dt} &= -KA \times u_{1}(t) \\
    \frac{du_{2}}{dt} &= KA \times u_{1}(t) - u_{2}(t) \times (KE + K12) + u_{3}(t) \times K21  \\
    \frac{du_{3}}{dt} &= K12 \times u_{2}(t) - K21 \times u_{3}(t)
\end{align}

where $c_{t} = u_{2}(t)/ \text{V2}$ with V2 representing the volume of distribution for drug in plasma circulation. The drift term in the relationship between $c_{t}$ and PD is represented 
by the following: 

\begin{align}
    \label{eqn:syn_data_ode_pd}
    \frac{du_{4}}{dt} = {KIN} - (KOUT * (1 - (Imax \times c_{t} / IC50 + c_t))) \times u_{4}(t).
\end{align}

\begin{figure}[ht]
\centering
\includegraphics[width=0.8\textwidth]{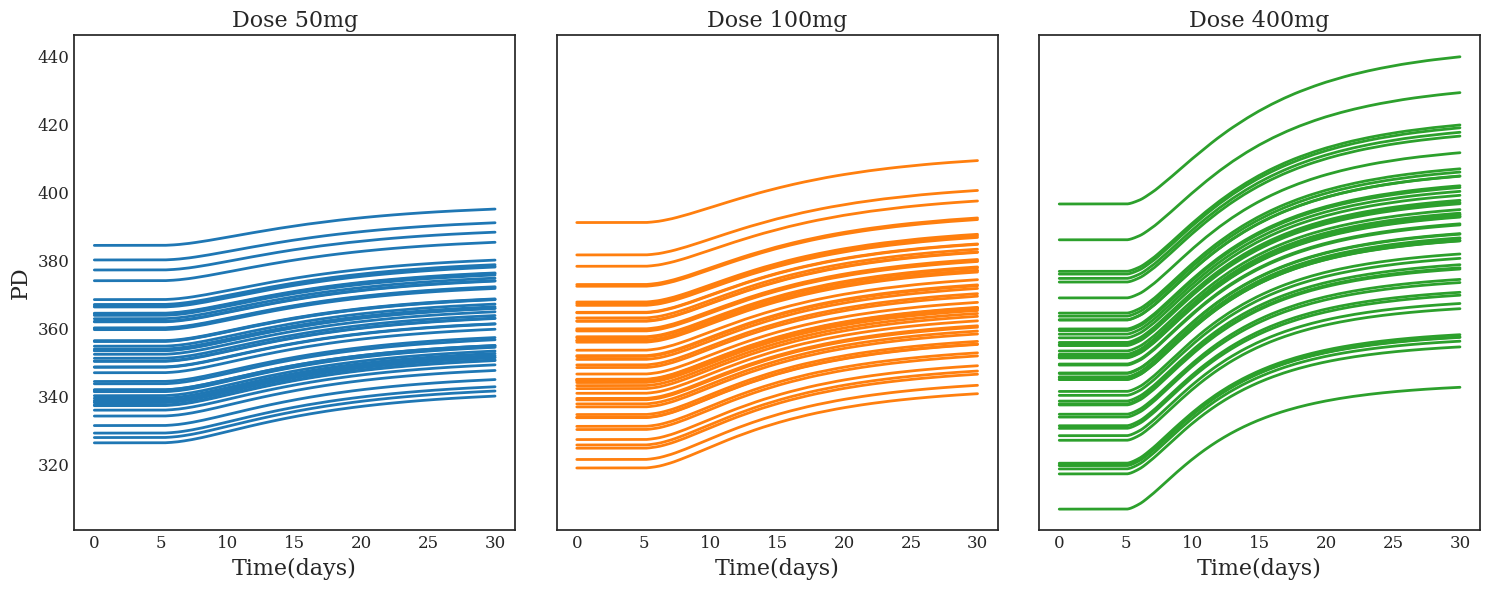}
\caption{Synthetic data trajectories without the diffusivity component under different simulated doses.}
\label{fig:Synthetic_data_ode}
\end{figure}

\begin{figure}[ht]
\centering
\includegraphics[width=0.8\textwidth]{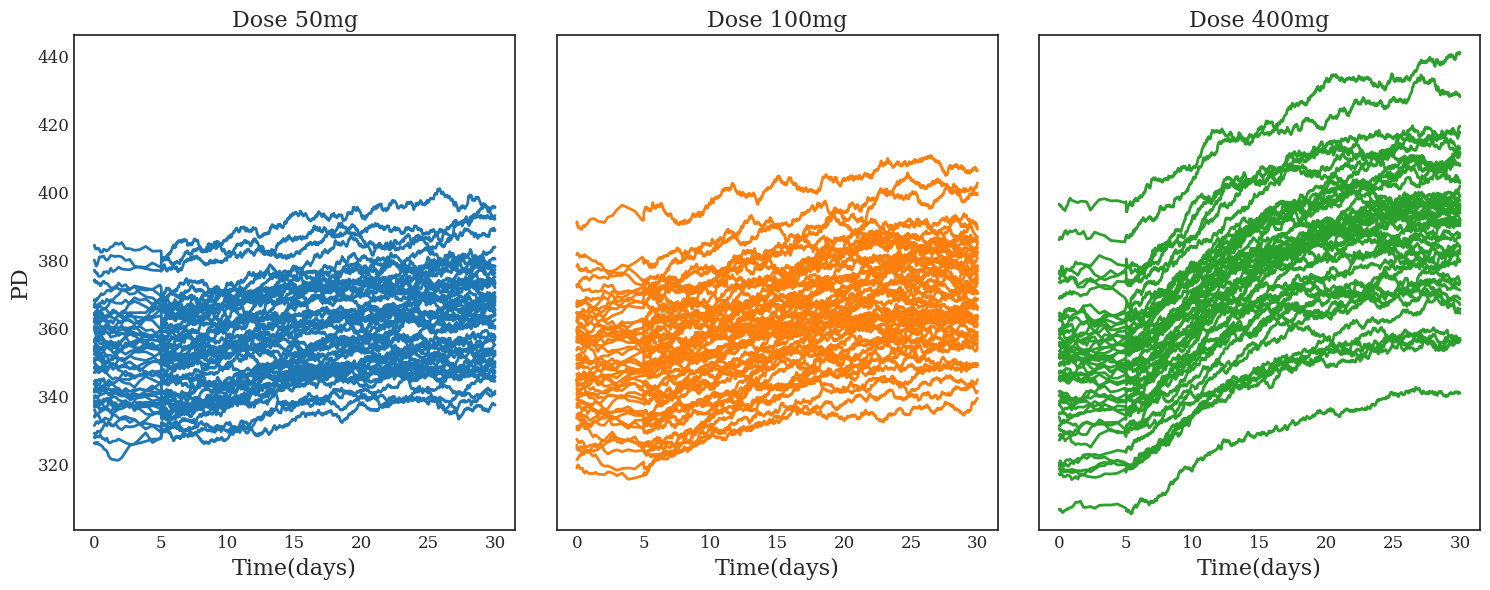}
\caption{Synthetic data trajectories under different doses.}
\label{fig:Synthetic_data_sde}
\end{figure}

Example trajectories of this system are shown in Figure \ref{fig:Synthetic_data_ode}.  The diffusion term in Equation \ref{eqn:sde} is described by the following 
$\beta u_{4} dW_t$, where $\beta$ was sampled from a log-normal distribution.
%
%
Examples of stochastic trajectories for $c_{t}$ are shown in Figure \ref{fig:Synthetic_data_sde}. 

In the current set of experiments, an equal number of patients were simulation for a range of doses (50, 100, 400 mg).  Dosing was set to begin at day 5 for all synthetic subjects with daily dosing; the PD sampling frequency is once per hour, over a period of 30 days.  

\subsection{Training methodology and optimization strategy }

The current model, including the numerical integration scheme which employs a  Euler-Maruyama solver, have been implemented in PyTorch. While a higher-order methods were not used in this current work, it remains open for future development based on specific needs. 

In model training, we leveraged vectorization rather than operating on a single value at a time whereby the model processes each time-step for each patient sequentially. In this way, the model operates at a patient level, concurrently processing all data points associated with a specific patient. This is feasible based on the observation that evaluating the loss function given in Equation \ref{eqn:Loss_Function} at each time-step is independent from other time instances. The vectorization strategy significantly enhances the training and inference performance. 

We trained the network for 100 epochs using the ADAM optimizer with learning rate 0.001 and batch size of 1. The overall training process takes around 140 seconds using one NVIDIA V100 GPU.

\end{document}